\documentclass[aps,prd,twocolumn,10pt,floatfix]{revtex4-2}
\usepackage{bm}
\usepackage{amsmath}
\usepackage{graphicx}
\usepackage{microtype}
\usepackage{latexsym}
\usepackage{amsfonts}
\usepackage{mathrsfs}
\usepackage{color}
\usepackage{hyperref}

\hypersetup{
	hidelinks,
}

\begin{document}

\title{Electro-Penrose Process in Charged BTZ Black Holes:\\
From Probe Limit to Thermodynamic Backreaction} 
	
\author{Hanjun Zou$^{1,2,3}$}
\email[Corresponding author:~]{zouhanjun24@mails.ucas.ac.cn}
\author{Ruo-Cen Wu$^{1,2,3}$}
\email[Corresponding author:~]{wuruocen24@mails.ucas.ac.cn}
\author{Meng Xu$^{1,2,3}$}
\email[Corresponding author:~]{xumeng241@mails.ucas.ac.cn}
\author{Zhoujian Cao$^{1,4}$}
\email[Corresponding author:~]{zjcao@bnu.edu.cn}
\author{Yun-Long Zhang$^{5,1}$}
\email[Corresponding author:~]{zhangyunlong@nao.cas.cn}
%\footnote{corresponding authors}
%[contact author:]

\affiliation{$^{1}$School of Fundamental Physics and Mathematical Sciences, Hangzhou Institute for Advanced Study, UCAS, Hangzhou 310024, China}
\affiliation{$^{2}$Taiji Laboratory for Gravitational Wave Universe (Beijing/Hangzhou), University of Chinese Academy of Sciences, Beijing 100049, China}
\affiliation{$^{3}$Institute of Theoretical Physics, Chinese Academy of Sciences, Beijing 100190, China}
\affiliation{$^{4}$School of Physics and Astronomy, Beijing Normal University, Beijing 100875, China}
\affiliation{$^{5}$National Astronomical Observatories, Chinese Academy of Sciences, Beijing 100101, China}

\begin{abstract}
We investigate the electro-Penrose process (EPP) in $(2+1)$-dimensional charged Ba\~nados-Teitelboim-Zanelli (BTZ) spacetime, from probe-limit particle dynamics to quasistatic thermodynamic backreaction. In the probe background, neutral particles have no exterior circular orbits, and charged circular motion requires a minimum repulsive coupling. We derive the maximum single-event energy at fixed decay data, including fragment masses and charge allocation, and check capture and collection at a finite receiver in anti-de Sitter (AdS) spacetime. Repetitive radial extraction on successive exact stationary backgrounds uses an adaptive incident specific charge. The continuum mass-charge path identifies the switch between incident turning-point and negative-energy stopping, supported by numerical checks. The prescribed sequences terminate at nonzero residual charge before exhausting the energy reservoir, providing a protocol-dependent third-law analogy. The consumed reservoir is divided between extracted energy and irreversible growth of irreducible mass, with proportions depending on the protocol.
\end{abstract}

\maketitle

\setlength{\abovedisplayskip}{6pt plus 2pt minus 2pt}
\setlength{\belowdisplayskip}{6pt plus 2pt minus 2pt}
\setlength{\abovedisplayshortskip}{4pt plus 2pt minus 2pt}
\setlength{\belowdisplayshortskip}{4pt plus 2pt minus 2pt}
\tableofcontents
\allowdisplaybreaks[1]

\section{Introduction}\label{intro}

\par The exploration of gravity in lower dimensions has provided valuable insights into the physics of black holes and quantum gravity. In this context, the $(2+1)$-dimensional BTZ black hole~\cite{Banados:1992wn, Banados:1992gq} serves as an important theoretical model. Characterized by a negative cosmological constant, the BTZ spacetime provides a framework that exhibits essential features of black holes, such as event horizons and thermodynamic properties, within the asymptotic structure of AdS geometry. Expanding upon the neutral case, the incorporation of electromagnetic fields introduces a purely radial electric field with a logarithmic gauge potential~\cite{Clement:1993kc, Martinez:1999qi}, which significantly modifies the interaction dynamics of the spacetime.

\par An established approach for probing the strong-field regime of black holes involves analyzing the dynamics of test particles and mechanisms of energy extraction. Originally proposed by Penrose for rotating Kerr black holes~\cite{Penrose:1969pc,Penrose:1971uk}, the Penrose process has been generalized to the EPP~\cite{Denardo:1973pyo,Bhat:1985,Wagh:1989zqa, Zaslavskii:2012ax,Tursunov:2019,Tursunov:2021jjf}, where energy extraction relies on electromagnetic fields rather than rotation. Particle dynamics in four-dimensional spacetimes such as Schwarzschild and Reissner-Nordstr\"om provide useful comparisons, but the AdS asymptotics and logarithmic electrostatic potential in $(2+1)$ dimensions lead to different conditions for orbital motion and particle transport. Specifically, the interplay between near-horizon gravitational pull and far-field AdS confinement significantly modifies the stability conditions for particle orbits compared to standard four-dimensional general relativity. Energy extraction therefore requires a consistent energy reference and an analysis of particle reachability alongside the local splitting kinematics.

\par Recently, the investigation of repetitive energy extraction processes and their corresponding thermodynamic backreaction has been conducted across various geometric backgrounds. Recent theoretical work has focused on exploring the iterative evolution and thermodynamic limits, including proposed dynamic third-law analogs for specific extraction protocols. These studies include Kerr-de Sitter~\cite{Wang:2025nfv}, accelerating Kerr~\cite{Zeng:2026egq}, Kerr-Taub-NUT black holes~\cite{Alloqulov:2026cyb}, rotating non-Kerr geometries~\cite{Zeng:2026jpn}, as well as genuine uncharged Kerr-Bertotti-Robinson geometries~\cite{Wu:2026nyg}. Related work on quantum-corrected Reissner-Nordstr\"om black holes examines single-event energy extraction and fragment escape~\cite{Chen:2026oxr}. More directly related to the present analysis, repetitive electric Penrose extraction in Reissner-Nordstr\"om spacetime has been studied with mass-charge updates and a proposed third-law analogy~\cite{Hu:2025bbc}. Recursive extraction in Reissner-Nordstr\"om-AdS spacetime has also been analyzed, with outgoing fragments returning for subsequent decays~\cite{Feiteira:2024awb,Feiteira:2026recursive}.

\par We analyze the EPP from probe-limit particle dynamics to quasistatic thermodynamic backreaction. Our aim is to examine the thermodynamic limits of the prescribed repetitive protocol by tracking irreducible-mass growth and determining its stopping mechanisms, and thereby assess a protocol-dependent third-law analogy. We derive an analytic mass-charge path in the infinitesimal-step limit and identify how the stopping mechanism changes. Section~\ref{sec2} gives the effective potential and necessary conditions for circular motion. Section~\ref{sec3} derives the maximum single-event energy at fixed decay data, including fragment masses and charge allocation, and checks capture and collection at a finite receiver. Section~\ref{sec4} analyzes the two stopping mechanisms of repetitive extraction and the partition of the consumed energy reservoir between extracted energy and irreversible mass growth. It examines how the decay radius, captured specific charge and incident energy affect the final efficiencies, with decreasing incident particle mass used to check numerical stability. Section~\ref{sec:dis} summarizes the main physical results and their scope.

\section{Background and equations of motion}\label{sec2}

We consider the framework of $(2+1)$-dimensional gravity with a negative cosmological constant, $\Lambda = -1/l^2$. The background geometry is described by the static, uncharged BTZ black hole spacetime, whose line element in the coordinates $(t, r, \phi)$ is expressed as~\cite{Banados:1992wn,Banados:1992gq}
\begin{equation}
    ds^2 = -f(r)dt^2 + f^{-1}(r)dr^2 + r^2d\phi^2,
\end{equation}
where the vacuum lapse function is given by
\begin{equation}
    f(r) = \frac{r^2}{l^2} - M,\label{eq:vacuum_lapse}
\end{equation}
parameterized by the black hole mass $M$ and the AdS curvature radius $l$. The event horizon is located at $r_h = l\sqrt{M}$, where $f(r_h) = 0$.

To incorporate electromagnetic interaction into this background, we introduce a static, radial test electric field generated by a central charge $Q$ located at the coordinate origin. By treating the charge in the probe limit, we neglect its backreaction on the background metric. In $(2+1)$ dimensions, the temporal component of the electromagnetic gauge potential $A_\mu$ takes a logarithmic form~\cite{Clement:1993kc,Martinez:1999qi,1982AnPhy.140..372D,Dunne:1998qy,carlip1998quantum}:
\begin{equation}
    A_t = Q \ln\left(\frac{r}{l}\right).\label{eq:potential}
\end{equation}
The corresponding Faraday tensor $F_{\alpha\beta} = \partial_\alpha A_\beta - \partial_\beta A_\alpha$ yields the non-zero component:
\begin{equation}
    F_{tr} = -F_{rt} = -\frac{Q}{r}.\label{eq:faraday}
\end{equation}

\subsection{Dynamics of a test particle } 

With the background geometry and electromagnetic field configuration established, we consider the dynamics of a test particle in this spacetime. The motion of a particle with rest mass $m$ and charge $q$ is governed by the Lorentz force equation~\cite{Cruz:1994ir,Cruz:2004ts,Tursunov:2021jjf}:
\begin{equation}
    \frac{du^\mu}{d\tau} + \Gamma^\mu{}_{\alpha\beta} u^\alpha u^\beta - \frac{q}{m} F^\mu_{\ \nu} u^\nu = 0,\label{eq:lorentz}
\end{equation}
where $u^\mu = dx^\mu/d\tau$ is the particle's unit timelike tangent, $\tau$ is the proper time, and $\Gamma^\mu{}_{\alpha\beta}$ are the Christoffel symbols associated with the BTZ metric.

The stationary and axisymmetric properties of the BTZ metric yield two Killing vector fields, $\partial_t$ and $\partial_\phi$, which correspond to the conserved specific energy (energy per unit mass) $\mathcal{E}$ and specific angular momentum (angular momentum per unit mass) $\mathcal{L}$, derived from the canonical momentum $P_\alpha = m u_\alpha + q A_\alpha$:
\begin{align}
    \frac{P_t}{m} &= -\mathcal{E} \equiv -\frac{E}{m} = u_t + \frac{qQ}{m} \ln\left(\frac{r}{l}\right),\label{eq:energy_integral} \\
    \frac{P_\phi}{m} &= \mathcal{L} \equiv \frac{L}{m} = u_\phi.\label{eq:angmom_integral}
\end{align}
We introduce a normalized charge-coupling parameter~\cite{Tursunov:2021jjf}:
\begin{align}
    \mathcal{Q} \equiv \frac{qQ}{m}.
\end{align}
The non-zero components of the equations of motion are given by
\begin{align}
    \frac{du^t}{d\tau} &= \frac{u^{r}}{f(r)}\left(\frac{\mathcal{Q}}{r} + u_{t}\frac{f'(r)}{f(r)}\right),\label{eq:eom_t} \\
    \frac{du^r}{d\tau} &= \frac{\mathcal{L}^{2}f(r)}{r^{3}}-\frac{\mathcal{Q}u_{t}}{r} + \frac{1}{2}\frac{f'(r)}{f(r)}\left((u^{r})^{2} - (u_{t})^{2}\right),\label{eq:eom_r} \\
    \frac{du^\phi}{d\tau} &= -\frac{2\mathcal{L}u^r}{r^3},\label{eq:eom_phi}
\end{align}
where $f'(r) = 2r/l^2$ and $u_t = -\mathcal{E} - \mathcal{Q}\ln(r/l)$. In the probe calculation we use the neutral BTZ lapse in Eq.~\eqref{eq:vacuum_lapse}. The conserved energy and angular momentum lead to the effective-potential description of radial motion below.

\subsection{Effective potential}

Using the normalization condition $u^\mu u_\mu=-k$, with $k=1$ for massive particles and $k=0$ for photons, and the conserved quantities, the radial motion decouples from the temporal and angular dynamics~\cite{Misner:1973prb,Chandrasekhar:1985kt}. For massive particles, $u^\mu=dx^\mu/d\tau$, $\mathcal E$, $\mathcal L$, and $\mathcal Q$ retain their specific definitions. For photons, $u^\mu=dx^\mu/d\lambda$ is an affinely parametrized null tangent, with $\mathcal E=-u_t$, $\mathcal L=u_\phi$, and $\mathcal Q=0$. The effective potential in the neutral BTZ metric with a test electric field is defined by the turning-point condition $\mathcal E=V_{\rm eff}(r)$:
\begin{equation}
    V_{\rm eff}(r)=-\mathcal Q\ln\left(\frac{r}{l}\right)+\sqrt{f(r)\left(k+\frac{\mathcal L^2}{r^2}\right)}.\label{eq:effective_potential}
\end{equation}

To describe coordinate kinematics we write $u^\mu=u^t(1,v,\Omega)$, with $v=dr/dt$ and $\Omega=d\phi/dt$. A static observer at finite radius instead has proper time $d\tau_{\rm stat}=\sqrt f\,dt$. Substituting this form into the normalization condition yields
\begin{equation}
    (u^t)^2\left(f^{-1}(r)v^2-f(r)+\Omega^2r^2\right)=-k.\label{eq:normalization_k}
\end{equation}
Solving for the angular velocity $\Omega$ gives
\begin{equation}
    \Omega=\pm\frac{1}{u_t r}\sqrt{u_t^2\left(f(r)-f^{-1}(r)v^2\right)-k f^2(r)}.\label{eq:angular_velocity}
\end{equation}
For a physical trajectory, the angular velocity $\Omega$ must be real, which requires the expression under the square root to be non-negative.

For photons ($k=0$), Eq.~\eqref{eq:angular_velocity} gives $r^2\Omega^2=f(r)-f^{-1}(r)v^2\le f(r)$ in the exterior region. Therefore, $\Omega$ is restricted to the interval $\Omega_-\le\Omega\le\Omega_+$, where the boundaries are
\begin{equation}
    \Omega_\pm=\pm\frac{\sqrt{f(r)}}{r},\label{eq:omega_pm}
\end{equation}
which corresponds to $k=0$ and $v=0$. However, vanishing radial velocity does not imply a circular geodesic. To examine whether a photon can remain at fixed radius, the null normalization condition gives
\begin{equation*}
    (u^r)^2=\mathcal E^2-\mathcal L^2\frac{f(r)}{r^2}.
\end{equation*}
Then the conserved quantities give
\begin{equation*}
    \frac{du^r}{d\lambda}=-\frac{\mathcal L^2}{2}\frac{d}{dr}\left(\frac{f(r)}{r^2}\right).
\end{equation*}
A circular null geodesic requires both $u^r=0$ and $du^r/d\lambda=0$. However, for $M>0$,
\begin{equation}
    \frac{d}{dr}\left(\frac{f(r)}{r^2}\right)=\frac{d}{dr}\left(\frac{1}{l^2}-\frac{M}{r^2}\right)=\frac{2M}{r^3}>0.
\end{equation}
So for $\mathcal L\ne0$ there is no finite-radius circular null geodesic in this background.

\begin{figure*}
    \centering
    \includegraphics[width=1\hsize]{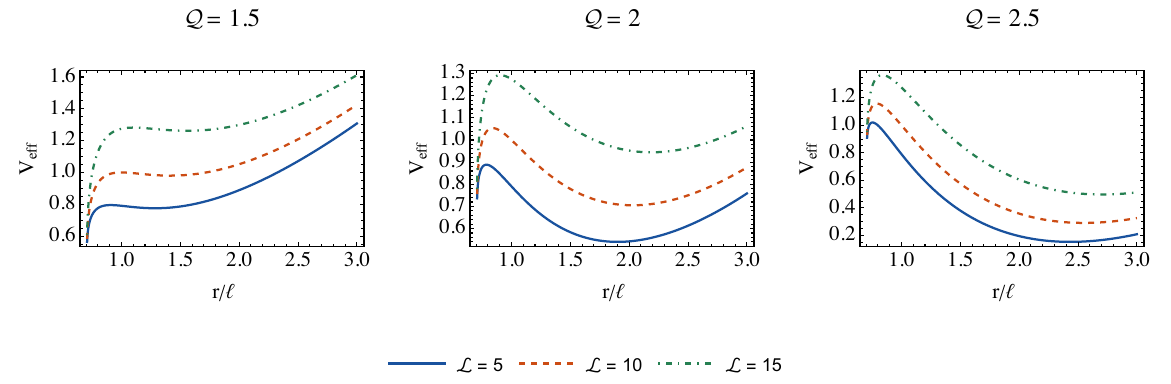}
    \caption{Radial dependence of the effective potential $V_{\mathrm{eff}}$ for a charged test particle in the neutral BTZ metric with a test electric field. The background black hole mass and the AdS radius are set to $M=0.5$ and $l=10$ respectively. The panels illustrate the influence of the $\mathcal L$ and the normalized charge-coupling parameter $\mathcal Q$. A local minimum, when present, gives stable circular motion; a local maximum is unstable. The separate reality and stability conditions are not replaced by the charge threshold.}\label{fig1}
\end{figure*}

For massive particles, setting $k=1$ in Eq.~\eqref{eq:effective_potential} recovers the effective potential governing a charged test particle. The sign of $\mathcal Q$ determines the nature of the electrostatic interaction. For $\mathcal Q>0$, the electrostatic repulsion opposes the gravitational attraction. Conversely, $\mathcal Q<0$ results in an attractive electrostatic interaction. The existence of circular orbits requires stationary points in the effective potential, determined by $\partial_r V_{\rm eff}(r)=0$~\cite{1972ApJ...178..347B}:
\begin{equation}
    J-Ml^2\mathcal L^2-r^4=0,\label{eq:polynomial_eq}
\end{equation}
where $J$ is defined as
\begin{equation}
    J=\mathcal Qlr\sqrt{\left(\mathcal L^2+r^2\right)\left(r^2-Ml^2\right)}.\label{eq:J_term}
\end{equation}
The two branches can be written compactly by introducing
\begin{equation}
    d_\pm=\mathcal Q\mp\sqrt{\mathcal Q^2-4M}.
\end{equation}
Their specific energies and squared angular momenta are
\begin{align}
    \mathcal E_\pm &=-\mathcal Q\ln(r/l)+\frac{2(r^2-Ml^2)}{l^2d_\pm},\\
    \mathcal L_\pm^2 &=\frac{r^2}{l^2}\frac{4r^2-2l^2\mathcal Qd_\pm}{d_\pm^2}.\label{eq:angular_momentum_roots}
\end{align}
For physical circular orbit, $\mathcal L$ must be real, which requires $\mathcal Q^2\ge4M$. Eq.~\eqref{eq:polynomial_eq} further requires $J>0$, hence $\mathcal Q>0$. These constraints give the
necessary lower bound:
\begin{equation}
    \mathcal Q\ge2\sqrt M.\label{eq:charge_condition}
\end{equation}
Within this probe background, neutral test particles have no exterior circular geodesics. For fixed nonzero $Q$, Eq.~\eqref{eq:charge_condition} constrains the particle's specific charge and selects repulsive electric coupling. The radial profile of the effective potential and the resulting stationary points corresponding to stable and unstable circular orbits are illustrated in Fig.~\ref{fig1} for representative values of $\mathcal{Q}$ and $\mathcal{L}$.

\section{Single-shot energy extraction}\label{sec3}

In the same probe background, we consider the local splitting of an incident particle 1 into fragments 2 and 3 at radius $r_{\rm d}$. Local conservation of energy, angular momentum, electric charge and radial momentum, together with the kinematic mass constraint, gives~\cite{Zaslavskii:2024zgh}
\begin{align}
    E_{1} &= E_{2} + E_{3} ,\quad L_{1} = L_{2} + L_{3}, \quad q_{1} = q_{2}  + q_{3},
    \\m_1 \dot{r_{1}} &= m_2 \dot{r_{2}} + m_3 \dot{r_{3}}, \quad m_1 \ge m_2 + m_3.\label{eq:conservation2}
\end{align}
Rest mass is not an additive conserved quantity in a general decay; the inequality expresses the available invariant mass. Here the overdot denotes differentiation with respect to the proper time $\tau$ of the corresponding particle. Using the definition of angular momentum $L_i = m_i g_{\phi\phi} u_i^\phi = m_i r_{\rm d}^2 u_i^\phi$, the conservation of angular momentum implies
\begin{equation}
    m_1 u_1^\phi = m_2 u_2^\phi + m_3 u_3^\phi.\label{eq:angular_momentum_uphi}
\end{equation}
We rewrite this equation using the kinematic relation $u_i^\phi = \Omega_i u_i^t$, where $\Omega_i = d\phi_i/dt$ is the coordinate angular velocity of the $i$-th particle. In the BTZ geometry, the temporal contravariant velocity is $u_i^t = -u_{ti}/f(r_{\rm d})$, and the covariant component is related to the energy and gauge potential via $m_i u_{ti} = -(E_i + q_i A_t)$. Substituting these expressions into the angular momentum equation, the metric factor $f(r_{\rm d})$ cancels out, yielding
\begin{equation}
    \Omega_1 (E_1 + q_1 A_t) = \Omega_2 (E_2 + q_2 A_t) + \Omega_3 (E_3 + q_3 A_t).\label{eq:omega_e_relation}
\end{equation}
Substituting the energy and charge conservation relations ($E_2 = E_1 - E_3$ and $q_2 = q_1 - q_3$) into the equation above, we obtain the energy of fragment 3:
\begin{equation}
    E_3 = \frac{\Omega_1 - \Omega_2}{\Omega_3 - \Omega_2} (E_1 + q_1 A_t) - q_3 A_t.\label{eq:E3_energy}
\end{equation}

\subsection{Maximum extraction efficiency}

To evaluate the optimal energy extraction, we assume the incident particle 1 is electrically neutral ($q_1=0$) and has initial specific energy $\mathcal E_1=1$ ($E_1=m_1$). This specifies a Killing energy, not rest at infinity in AdS. We take the particle to be prepared locally at its radial turning point; a neutral particle cannot arrive at this outer turning point from a larger radius. The preparation cost is not included. At the turning point ($u^r=0$), the velocity normalization condition yields the angular velocity of particle 1 at the splitting radius $r_{\rm d}$:
\begin{equation}
    \Omega_1 = \frac{1}{r_{\rm d}}\sqrt{\left(\frac{r_{\rm d}^2}{l^2} - M\right)\left(1 + M - \frac{r_{\rm d}^2}{l^2}\right)}.\label{eq:Omega1_final}
\end{equation}
To assess the efficiency of this energy extraction, we maximize the energy acquired by particle 3 at fixed background, splitting radius, parent state, daughter masses and charge allocation. Only the emission direction in the parent rest frame is varied. Write $f_{\rm d}=f(r_{\rm d})$, $\tilde\mu_i=m_i/m_1$, and
\begin{equation}
\begin{aligned}
    B_3&=1+\tilde\mu_3^2-\tilde\mu_2^2,\\
    \Delta&=[1-(\tilde\mu_2+\tilde\mu_3)^2][1-(\tilde\mu_2-\tilde\mu_3)^2].
\end{aligned}
\label{eq:single_event_masses}
\end{equation}
Local energy-momentum conservation gives the energy and momentum magnitude of particle 3 in the parent rest frame as $\varepsilon_3^*=m_1B_3/2$ and $p^*=m_1\sqrt\Delta/2$. The parent has speed $\beta=\sqrt{1-f_{\rm d}}$ and Lorentz factor $\gamma=1/\sqrt{f_{\rm d}}$ relative to a static observer at the splitting point. If $\theta$ is the emission angle measured from the parent's tangential direction, that observer measures
\begin{equation}
\begin{aligned}
    \varepsilon_3&=\gamma(\varepsilon_3^*+\beta p^*\cos\theta),\\E_3&=\sqrt{f_{\rm d}}\,\varepsilon_3-q_3A_t(r_{\rm d}).
\end{aligned}
\label{eq:single_event_boost}
\end{equation}
Since $\cos\theta\le1$, the maximum is attained for forward emission~($\theta=0$)~\cite{Zaslavskii:2024zgh}:
\begin{equation}
    E_{3,\max}=\frac{m_1}{2}\left[B_3+\sqrt{(1-f_{\rm d})\Delta}\right]-q_3A_t(r_{\rm d}).\label{eq:finite_mass_max}
\end{equation}
This result includes finite daughter masses and obeys the full local conservation laws. For the maximizing choice $\theta=0$, both daughter momenta are tangential and their instantaneous radial momenta vanish. Capture and collection must still be checked below.

In the ultrarelativistic limit $\tilde\mu_2,\tilde\mu_3\to0$, these tangential momenta approach opposite null directions. Using the limiting values derived in Eq.~(\ref{eq:omega_pm}), we have
\begin{align}
    \Omega_3 &=-\Omega_2 = \frac{1}{r_{\rm d}} \sqrt{\frac{r_{\rm d}^2}{l^2} - M}.\label{eq:Omega3_max}
\end{align}
Substituting these limiting angular velocities into Eq.~(\ref{eq:E3_energy}), the limiting maximum energy of particle 3 is
\begin{equation}
\begin{aligned}
    E_{3,\max}^{\rm UR} &= \frac{m_1}{2} \left( 1 + \sqrt{1 + M - \frac{r_{\rm d}^2}{l^2}} \right)\\&\quad - q_3 Q \ln(r_{\rm d}/l).
\end{aligned}
\label{eq:E3_max_BTZ}
\end{equation}
To parameterize the allowed decay radius, we introduce a dimensionless parameter $\alpha\in(0,1]$ defined by $r_{\rm d}^2=l^2(M+\alpha)$, so $f_{\rm d}=\alpha$. This parameterization places the splitting outside the probe horizon ($r_{\rm d}>l\sqrt M$) while maintaining a real value for the square root in Eq.~(\ref{eq:E3_max_BTZ}). The limiting maximum energy extraction efficiency can be expressed in terms of the initial particle's rest mass $m_1$ and the fragment's charge $q_3$ as
\begin{equation}
    \frac{E_{3,\max}^{\rm UR}}{E_1} = \frac{1}{2} + \frac{1}{2}\sqrt{1 - \alpha} - \frac{q_3}{2 m_1}Q \ln(M + \alpha).\label{eq:efficiency_2d}
\end{equation}
Here $E_3/E_1$ is the gross energy ratio; the net extraction efficiency is $E_3/E_1-1$. Eq.~\eqref{eq:efficiency_2d} is the ultrarelativistic limit of the fixed-mass maximum in Eq.~\eqref{eq:finite_mass_max}, not a maximum over arbitrary daughter masses. For equal masses $\tilde\mu_2=\tilde\mu_3=\mu$, its square-root term is replaced by $\sqrt{(1-\alpha)(1-4\mu^2)}$, so the limiting value is approached from below as $\mu\to0$. No electrically charged photons are assumed.
\begin{figure*}
    \centering
    \includegraphics[width=1\hsize]{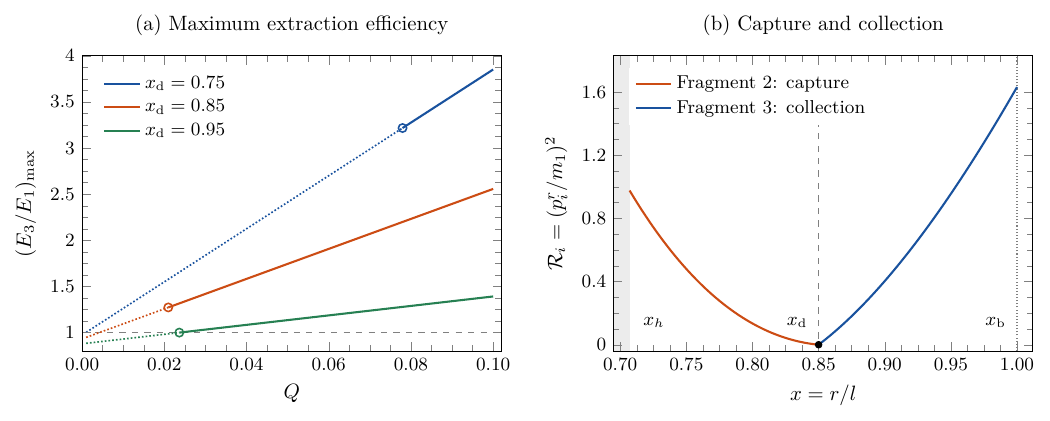}
    \caption{Maximum single-event energy extraction in the probe model for $M=0.5$, $l=10$, $\tilde\mu_2=\tilde\mu_3=0.1$ and $q_3/m_1=100$. (a) Fixed-mass maximum versus the test-field charge $Q$ at three splitting radii. Solid curves satisfy negative-energy capture and collection at $x_{\rm b}=1$; dotted curves give the local maximum where at least one requirement fails for the maximizing direction. The solid-to-dotted transition is set by the outward-launch condition at $x_{\rm d}=0.75$ and $0.85$, and by the negative-energy condition at $x_{\rm d}=0.95$. The corresponding lower thresholds are $Q_{\rm crit}\simeq0.07795375$, $0.02096707$ and $0.02365184$ respectively; both thresholds are listed in Table~\ref{tab:single_event_thresholds}. Open circles mark excluded marginal configurations. (b) Allowed radial motion of both fragments at $Q=0.05$, $x_{\rm d}=0.85$, with $x_h=\sqrt M$ and $E_{3,\max}/E_1=1.74456687$.}\label{fig3}
\end{figure*}

The quantitative behavior of the finite-mass maximum is illustrated in Fig.~\ref{fig3}. For the sign choice $q_3Q>0$, obtaining $E_3/E_1>1$ requires $\ln(M+\alpha)<0$, hence $M<1$. This is a necessary condition in the stated gauge and fixed-parent setting, not a universal mass bound for the EPP of BTZ. To identify when the maximum is an extractable energy, we require particle 2 to enter the horizon with $E_2<0$ and particle 3 to reach a finite receiver at $x_{\rm b}=1$, where $x=r/l$ and $A_t=0$. These conditions can be checked with the full radial equation, retaining the daughters' angular momenta. Define $e_i=E_i/m_1$, $\kappa_i=q_iQ/m_1$, $j_i=L_i/(m_1l)$ and the mechanical momentum $p_i^\mu=m_i u_i^\mu$, then we have
\begin{align}
    K_i(x)&=e_i+\kappa_i\ln x,\nonumber\\
    \mathcal R_i(x)&\equiv\left(\frac{p_i^r}{m_1}\right)^2=K_i^2-(x^2-M)\left(\tilde\mu_i^2+\frac{j_i^2}{x^2}\right),\label{eq:single_event_radial}\\
    \frac{x}{2}\mathcal R_i'(x)&=\kappa_iK_i-x^2\tilde\mu_i^2-\frac{Mj_i^2}{x^2}.\label{eq:single_event_slope}
\end{align}
Future-directed motion requires $K_i>0$. At the maximizing split, $e_3=E_{3,\max}/m_1$, $e_2=1-e_3$, $\kappa_2=-\kappa_3$, and
\begin{align}
    j_3&=\frac{x_{\rm d}}{2\sqrt\alpha}\left[\sqrt{1-\alpha}\,B_3+\sqrt\Delta\right],\nonumber\\
    j_2&=x_{\rm d}\sqrt{\frac{1-\alpha}{\alpha}}-j_3,\label{eq:single_event_angular}
\end{align}
which give $\mathcal R_i(x_{\rm d})=0$. Writing
\begin{equation}
    K_{3{\rm d}}\equiv K_3(x_{\rm d})=\frac12\left[B_3+\sqrt{(1-\alpha)\Delta}\right],\label{eq:single_event_K3d}
\end{equation}
we also have $K_2(x_{\rm d})=1-K_{3{\rm d}}>0$ for the massive daughters considered here. In the probe model, $K_{3{\rm d}}$ and $j_i$ are independent of $Q$ at fixed $M$, $x_{\rm d}$ and daughter masses.

For $\kappa_3>0$, $K_2$ increases inward and $\mathcal R_2'<0$ between the splitting point and the probe horizon, ensuring capture. Particle 3 requires $\mathcal R_3'(x_{\rm d})>0$ and $K_3,\mathcal R_3>0$ throughout $x_{\rm d}<x\le1$. If these conditions and $e_2<0$ hold, the local maximum is attained within the allowed extraction processes at the specified fixed data. Otherwise it is only a local upper bound; a maximum over the remaining allowed directions would require a separate optimization.

The two restrictions give distinct charge thresholds. Since
\begin{equation}
    e_2=1-e_3=1-K_{3{\rm d}}+\kappa_3\ln x_{\rm d},
\end{equation}
where $K_{3{\rm d}}=K_3(x_{\rm d})$. The negative-energy condition at $x_{\rm d}<1$ requires
\begin{equation}
    \kappa_3>\kappa_{\rm neg}\equiv\frac{1-K_{3{\rm d}}}{-\ln x_{\rm d}}.\label{eq:single_event_negative_threshold}
\end{equation}
Likewise, Eq.~\eqref{eq:single_event_slope} gives the outward-launch condition
\begin{equation}
    \kappa_3>\kappa_{\rm out}\equiv\frac{x_{\rm d}^2\tilde\mu_3^2+Mj_3^2/x_{\rm d}^2}{K_{3{\rm d}}}.\label{eq:single_event_outward_threshold}
\end{equation}
To check that outward launch leads to collection, differentiate Eq.~\eqref{eq:single_event_slope}:
\begin{equation}
    \frac{d}{dx}\left(\frac{x}{2}\mathcal R_3'\right)=\frac{\kappa_3^2}{x}-2x\tilde\mu_3^2+\frac{2Mj_3^2}{x^3}.\label{eq:single_event_outward_monotonicity}
\end{equation}
For $M>0$, $x\le1$ and $\kappa_3^2>2\tilde\mu_3^2$, this derivative is positive. Thus $\mathcal R_3'(x_{\rm d})>0$ ensures $\mathcal R_3'>0$ and $\mathcal R_3>0$ up to the receiver. Moreover, $K_3(x)=K_{3{\rm d}}+\kappa_3\ln(x/x_{\rm d})>0$ along the outward path.

For $q_3>0$, define the sufficient charge threshold
\begin{equation}
\begin{aligned}
    Q_{\rm neg}&=\frac{m_1}{q_3}\kappa_{\rm neg},\qquad Q_{\rm out}=\frac{m_1}{q_3}\kappa_{\rm out},\\
    Q_{\rm crit}&=\max\left\{Q_{\rm neg},Q_{\rm out},\frac{m_1}{q_3}\sqrt{2}\,\tilde\mu_3\right\}.
\end{aligned}
\label{eq:single_event_charge_threshold}
\end{equation}
Then $Q>Q_{\rm crit}$ ensures negative captured energy, outward launch and the monotonicity condition $\kappa_3^2>2\tilde\mu_3^2$, guaranteeing capture and collection for the maximizing split. For the parameters in Fig.~\ref{fig3}, the monotonicity threshold is $(m_1/q_3)\sqrt{2}\,\tilde\mu_3\simeq0.00141421$, below $\max(Q_{\rm neg},Q_{\rm out})$ at all three plotted radii, as shown in Table~\ref{tab:single_event_thresholds}. The table also identifies which condition determines the start of each solid curve. All solid portions in Fig.~\ref{fig3} satisfy $Q>Q_{\rm crit}$.

\begin{table}[t]
    \centering
    \setlength{\tabcolsep}{4.5pt}
    \renewcommand{\arraystretch}{1.08}
    \caption{Negative-energy and outward-launch thresholds for the maximizing split in Fig.~\ref{fig3}(a), with $M=0.5$, $\tilde\mu_2=\tilde\mu_3=0.1$ and $q_3/m_1=100$. The larger threshold determines the start of each solid curve; accepted configurations require $Q>Q_{\rm crit}$.}\label{tab:single_event_thresholds}
    \begin{tabular}{c c c c}
        \hline\hline
        $x_{\rm d}$
        & $Q_{\rm neg}$
        & $Q_{\rm out}$
        & Limiting condition \\
        \hline
        $0.75$ & $0.000892$ & $0.0780$ & Outward launch \\
        $0.85$ & $0.00419$  & $0.0210$ & Outward launch \\
        $0.95$ & $0.0237$   & $0.0110$ & Negative energy \\
        \hline\hline
    \end{tabular}
\end{table}

As an explicit example, with $M=0.5$, $Q=0.05$, $x_{\rm d}=0.85$, $\tilde\mu_2=\tilde\mu_3=0.1$ and $q_3/m_1=100$, we obtain $E_{3,\max}/E_1=1.74456687$ and $E_2/E_1=-0.74456687$. Both thresholds are satisfied, and the radial functions in Fig.~\ref{fig3}(b) show capture and collection for this maximizing configuration. This establishes a maximum for a locally prepared parent in the probe model. It is separate from the incident transport protocol on exact charged backgrounds in Section~\ref{sec4}.

\section{Repetitive extraction and backreaction}\label{sec4}

The analysis in the preceding sections, conducted under the test-field approximation, shows that the $(2+1)$-dimensional weakly charged BTZ black hole allows for efficient energy extraction through the EPP. In a repetitive sequence, the black hole's electric charge and mass evolve through successive absorptions, so the background geometry and particle trajectories must be updated at each stage.

To maintain theoretical consistency in investigating the long-term evolution and thermodynamic limits of the system, we transition from the probe limit to a quasistatic framework that incorporates the electromagnetic contribution to the background geometry. We now employ the exact charged BTZ metric to describe each stationary stage, while treating the absorption-induced changes of its parameters in the test-particle approximation. In $(2+1)$ dimensions, the coupling between the charge $Q$ and the metric is important, as the logarithmic potential not only governs the particle dynamics but also modifies the event horizon and the irreducible mass.

The exact line element for the charged BTZ black hole is given by~\cite{Martinez:1999qi,Du:2023nkr}
\begin{equation}
    ds^2=-f_Q(r)dt^2+\frac{dr^2}{f_Q(r)}+r^2d\phi^2,\label{eq:exact_metric}
\end{equation}
where the charged lapse function including the electromagnetic backreaction on the stationary geometry is defined as
\begin{equation}
    f_Q(r)=-M+\frac{r^2}{l^2}-\frac{Q^2}{2}\ln\frac r l.\label{eq:exact_f}
\end{equation}
We use the dimensionless radius and lapse
\begin{equation}
    x=r/l,\qquad F(x;M,Q)=x^2-M-\frac{Q^2}{2}\ln x.
\end{equation}
By updating $M$ and $Q$ after each iteration~\cite{Ruffini:2024dwq,Hu:2025bbc}, this framework allows us to track the irreversible growth of the irreducible mass and establish a third-law analog on the repetitive EPP. We also investigate the stopping mechanisms and extraction limits of the prescribed repetitive protocol, to which the third-law analogy is restricted. Finally, the existence of an event horizon is a prerequisite for the EPP. For the charged BTZ lapse function Eq.~\eqref{eq:exact_f}, to avoid a naked singularity, the mass $M$ and charge $Q$ must satisfy the regularity condition
\begin{equation}
    M\ge M_{\rm ext}(Q)\equiv\frac{Q^2}{4}-\frac{Q^2}{2}\ln\frac{|Q|}{2}.\label{eq:horizon_condition}
\end{equation}
For $Q\ne0$, $f_Q'(r)=2r/l^2-Q^2/(2r)$ vanishes only at $r_0=l|Q|/2$, where $f_Q''(r_0)=4/l^2>0$. Since $f_Q\to+\infty$ as $r\to0^+$ and $r\to\infty$, a horizon exists precisely when this minimum satisfies $f_Q(r_0)\le0$, giving the condition above. Strict inequality gives an inner and an outer horizon; equality gives the extremal, degenerate horizon at $r_0$. In our subsequent analysis, all parameters are maintained within this regime. Throughout this section, $r_+$ denotes the largest positive root of $f_Q(r)=0$, and we define $x_+=r_+/l$. Our numerical protocol uses $0<M<1$, $0<Q<2$ and $x_+<x_{\rm d}<1$. Since $M_{\rm ext}'(Q)=-Q\ln(Q/2)>0$ for $0<Q<2$, this branch has a unique extremal charge.

\subsection{Radial extraction protocol}\label{sec:protocol}

We keep $l$, the time coordinate $t$, and the logarithmic reference scale fixed, and use the Killing energy $E=-P_t$ in the gauge $A_t=Q\ln x$. The source and receiver are at $x_{\rm b}=1$, where $A_t=0$ and a static observer measures $E_{\rm loc}=E/\sqrt{1-M}$. Within each stationary stage, incident and collected energies acquire the same redshift factor, so their ratio equals the Killing-energy ratio. Across stages we sum Killing energies in this common convention. Particle 3 need only return to the finite receiver; finite-energy massive particles do not escape to the AdS boundary. We use $c=8G=1$ and the electromagnetic normalization of the specified $f_Q$ and $A_t$. Particle masses, energies and charges are expressed in the same geometric units as the black-hole parameters, so absorption gives $\delta M=E_2$ and $\delta Q=q_2$.

In each sequence we fix
\begin{equation}
\begin{gathered}
    \{m_1,\mathcal E_1,x_{\rm d},\eta,\hat q_2,l\},\\
    \hat q_i=q_i/m_i,\quad\eta=\tilde\mu_3/\tilde\mu_2,
\end{gathered}
\label{eq:fixed_data}
\end{equation}
where $\tilde\mu_i=m_i/m_1$ and $\eta=m_3/m_2$. The incident mass is fixed relative to the \emph{initial} mass, $m_1=M_0/k$ where $k$ is a dimensionless constant, and is not rescaled after each absorption. We choose an idealized threshold splitting with all three particles instantaneously at rest radially and with zero angular momentum at $x_{\rm d}$. Local energy-momentum conservation then requires $m_1=m_2+m_3$.

For radial motion on stage $n$, we define
\begin{align}
    V_{i,n}^\pm(x)&=-\hat q_{i,n}Q_n\ln x\pm\sqrt{F_n(x)},\\
    R_{i,n}(x)&=(u_i^r)^2=(\mathcal E_{i,n}+\hat q_{i,n}Q_n\ln x)^2-F_n(x),\label{eq:radial_exact}\\
    F_n(x)&=F(x;M_n,Q_n).
\end{align}
Future-directed exterior motion requires $\mathcal E_{i,n}+\hat q_{i,n}Q_n\ln x>0$. Together with $R_{i,n}\ge0$, this selects $\mathcal E_{i,n}\ge V_{i,n}^+$. Writing $S_n=\sqrt{F_n(x_{\rm d})}$, the splitting data are
\begin{align}
    \tilde\mu_2=&\frac1{1+\eta},\quad \tilde\mu_3=\frac\eta{1+\eta},\\
    \hat q_{1,n}&=\frac{S_n-\mathcal E_1}{Q_n\ln x_{\rm d}},\label{eq:adaptive_charge}\\
    \hat q_{3,n}&=\frac{\hat q_{1,n}-\tilde\mu_2\hat q_2}{\tilde\mu_3},\\
    \mathcal E_{2,n}&=-\hat q_2Q_n\ln x_{\rm d}+S_n,\label{eq:negative_energy}\\
    \mathcal E_{3,n}&=\frac{\mathcal E_1-\tilde\mu_2\mathcal E_{2,n}}{\tilde\mu_3}.
\end{align}
These relations enforce charge and local energy-momentum conservation, as well as $R_{i,n}(x_{\rm d})=0$ for all three particles.\label{par:adaptive} Only the new incident and outgoing particles have their specific charges adjusted between stages. Each particle's charge is constant along its own trajectory.

When particle 2 is absorbed, its energy and charge update the black hole. With $n$ counting completed absorptions, the recurrence is~\cite{Hu:2025bbc}
\begin{align}
    M_{n+1}&=M_n+m_1\tilde\mu_2\mathcal E_{2,n},\label{eq:update_mass}\\
    Q_{n+1}&=Q_n+m_1\tilde\mu_2\hat q_2.\label{eq:update}
\end{align}

\subsection{Reachability and the two limiting conditions}\label{sec:conditions}

Within the admissible exterior and source domain, extraction requires the incident particle to reach the prescribed decay point and the captured fragment to have negative energy. Failure of either condition stops the sequence. For the incident particle from larger radii, an ordinary inner turning point satisfies $\mathcal E_1=V_1^+(x_{\rm d})$ and
\begin{align}
    D_{1,n}=\left.\frac{dV_{1,n}^+}{dx}\right|_{x_{\rm d}}=\frac1{x_{\rm d}}\left[-\hat q_{1,n}Q_n+\frac{x_{\rm d}^2-Q_n^2/4}{S_n}\right]<0.\label{eq:turn_slope}
\end{align}
The limiting case occurs when the turning point coincides with the peak of the effective potential, where $\mathcal E_1=V_1^+(x_{\rm d})$ and $D_{1,n}=0$. Indeed, at the turning point $R'_{1,n}=-2S_nD_{1,n}$, hence $D_{1,n}<0$ makes the immediately exterior side accessible.

\begin{figure*}[t]
\centering
\includegraphics[width=.48\textwidth]{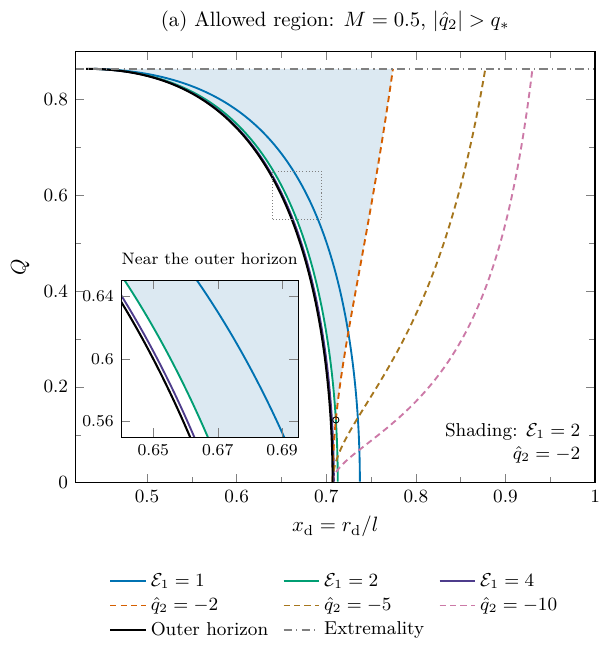}\hfill
\includegraphics[width=.48\textwidth]{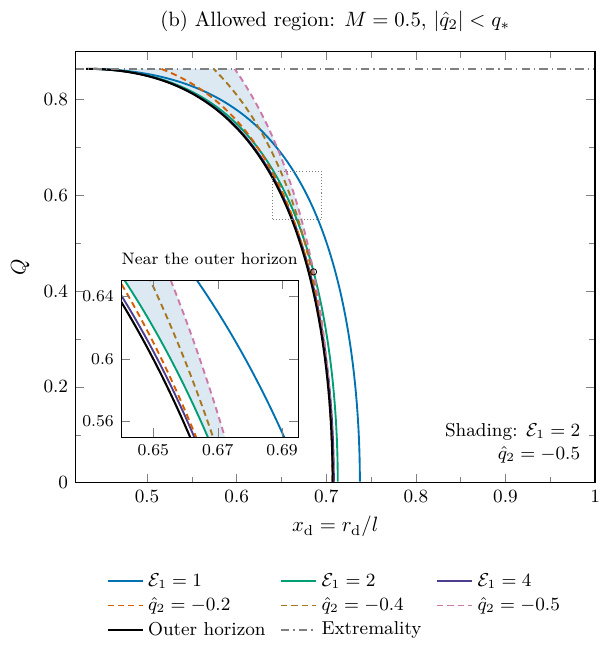}
    \caption{Constraints in the $(x_{\rm d},Q)$ plane at $M=0.5$. Solid curves satisfy the simultaneous incident critical conditions, with $\hat q_1$ determined by Eq.~\eqref{eq:adaptive_charge}; dashed curves give $\mathcal E_2=0$. Panels (a,b) correspond to $|\hat q_2|>q_*$ and $|\hat q_2|<q_*$. The black curve is the outer horizon; the gray line is extremality on the $0<Q<2$ branch. Shading refers to $(\mathcal E_1,\hat q_2)=(2,-2)$ in (a) and $(2,-0.5)$ in (b), subject to the exterior and source conditions. Open circles mark joint marginal boundaries. Insets enlarge the dotted rectangles. These are fixed-mass slices; the repetitive sequence changes both $M$ and $Q$.}\label{fig:allowed_region}
\end{figure*}

To get the first condition we substitute Eq.~\eqref{eq:adaptive_charge} into the limiting condition $D_{1,n}=0$ and suppressing the stage index $n$ yields
\begin{equation}
    F_{\rm d}-\mathcal E_1\sqrt{F_{\rm d}}-(x_{\rm d}^2-Q^2/4)\ln x_{\rm d}=0,\label{eq:critical_consistent}
\end{equation}
here $F_{\rm d}=F(x_{\rm d};M,Q)$. The solid curves in Fig.~\ref{fig:allowed_region} mark the limiting incident barrier boundaries for $\mathcal E_1=1,2,4$ at $M=0.5$. These curves are identical in the two panels.

The second condition requires the captured fragment must satisfy $\mathcal E_2<0$, with equality defining the limiting boundary. For $\hat q_2<0$, $Q>0$, $0<x_{\rm d}<1$ and $F_{\rm d}>0$, Eq.~\eqref{eq:negative_energy} is equivalent to
\begin{equation}
    A(x_{\rm d})Q^2>x_{\rm d}^2-M,
\end{equation}
where $A(x)=\ln x(\hat q_2^2\ln x+1/2)$. For $0<M<1$, define $x_0=\sqrt M$, $x_e=\exp[-1/(2\hat q_2^2)]$ and $q_*=\sqrt{-1/\ln M}$. Since $A>0$ for $x<x_e$ and $A<0$ for $x_e<x<1$, the real positive-charge boundary
\[Q_{\rm neg}(x_{\rm d})=\sqrt{\frac{x_{\rm d}^2-M}{A(x_{\rm d})}}\]
gives
\[\begin{cases}
    Q>Q_{\rm neg}, & x_0<x_{\rm d}<x_e,\quad |\hat q_2|>q_*,\\
    Q<Q_{\rm neg}, & x_e<x_{\rm d}<x_0,\quad |\hat q_2|<q_*.
\end{cases}\]
Negative-energy states thus lie above or below the curve respectively. For $x_{\rm d}<\min(x_0,x_e)$ the inequality holds automatically, whereas for $x_{\rm d}>\max(x_0,x_e)$ it cannot hold. At $x_{\rm d}=x_e$ it reduces to $M>x_e^2$ while $|\hat q_2|=q_*$ the boundary is $x_{\rm d}=x_0=x_e$. All these regions must also lie outside the actual charged outer horizon $x_+$ and satisfy the black-hole existence and reachability conditions. These two regimes are illustrated in Fig.~\ref{fig:allowed_region}(a) and (b). The dashed curves mark $\mathcal E_2=0$. Along the corresponding boundary branches, negative-energy states lie above the curves in panel (a) and below them in panel (b). 

For an initially admissible state on the branch $Q<Q_{\rm neg}$, $x_e<x_{\rm d}<x_0$ and $|\hat q_2|<q_*$, the captured energy remains strictly negative along the continuum path of the prescribed protocol with fixed $x_{\rm d}$ and $\hat q_2$. This branch therefore cannot terminate through loss of negative-energy capture. Among the two limiting mechanisms, only the incident turning-point condition can set the extraction limit, provided the other applicability conditions remain satisfied. Appendix~\ref{app:continuum} proves this statement and gives a sufficient condition for its validity at finite step.

Reachability must also hold along the path to the finite source. In our exterior radial sector, this can be checked analytically. Set $N=x^2-Q^2/4$, $w=N/\sqrt F$, and $H=2x^2F-N^2$. Since the outer horizon satisfies $x_+\ge |Q|/2$, we have $N=x^2-Q^2/4>0$ throughout the exterior region $x>x_+$. Then
\begin{equation}
\begin{aligned}
    V_i^{+\prime}&=\frac{w-\hat q_iQ}{x},\\
    w'&=\frac{H}{xF^{3/2}},\qquad H'=4xF>0.
\end{aligned}
\end{equation}
Thus $w$ has at most one minimum. Starting with $D_1<0$, $V_1^+$ can only decrease and then increase on $[x_{\rm d},1]$. The endpoint condition is
\begin{equation}
    \mathcal E_1\ge V_1^+(1)=\sqrt{1-M}.\label{eq:source_gate}
\end{equation}
Since the turning-point condition already fixes $\mathcal E_1=V_1^+(x_{\rm d})$, the endpoint condition therefore ensures $\mathcal E_1\ge V_1^+(x)$ throughout this interval. For $\hat q_2<0$ and $Q>0$, $V_2^{+\prime}>0$ outside the horizon, so particle 2 can fall inward. Outside the outer horizon $N>0$, so $D_1<0$ implies $\hat q_1Q>N/S>0$ and hence $\hat q_1>0$. It follows that $\hat q_3-\hat q_1=(\hat q_1-\hat q_2)/\eta>0$. Define $D_3= V_3^{+\prime}(x_{\rm d})$ then
\begin{equation*}
    D_3=D_1-\frac{(\hat q_3-\hat q_1)Q}{x_{\rm d}}<D_1<0.
\end{equation*}
The immediately exterior region is therefore accessible. Along the full outward path,
\begin{equation}
    \mathcal E_3-V_3^+(x)=\mathcal E_1-V_1^+(x)+(\hat q_3-\hat q_1)Q\ln(x/x_{\rm d})\ge0
\end{equation}
on $[x_{\rm d},1]$. It is strict for $x_{\rm d}<x\le1$, so particle 3 remains future-directed, has no further turning point and reaches the receiver at $x=1$. This establishes full-path reachability.

Given the incident specific energy $\mathcal E_1$ and the captured fragment's specific charge $\hat q_2$, these conditions constrain both $x_{\rm d}$ and $Q$. Fig.~\ref{fig:allowed_region} illustrates the allowed domain at fixed black-hole mass. The solid curves give the incident barrier boundary, while the dashed curves give the negative-energy boundary. The allowed region is the intersection of the two inequalities with the exterior and source conditions. At fixed $(M,Q,x_{\rm d})$, comparison between adaptive-charge experiments gives
\begin{equation}
\begin{aligned}
    \frac{\partial D_1}{\partial\mathcal E_1}&=\frac{1}{x_{\rm d}\ln x_{\rm d}}<0,\\
    \frac{\partial\mathcal E_2}{\partial|\hat q_2|}&=Q\ln x_{\rm d}<0.
\end{aligned}
\end{equation}
Increasing either parameter therefore relaxes its corresponding gate, but an allowed interval requires a nonempty intersection of all conditions. Intersections of the two limiting curves are joint marginal boundaries at fixed mass. A repetitive sequence changes both mass and charge; provided the exterior and source conditions remain satisfied, its continuous-limit endpoint is the first limiting boundary encountered along the evolving path. The finite-step stopping rule is specified in Section~\ref{sec:iteration}.

\subsection{Thermodynamic evolution and energy accounting}\label{sec:iteration}

Each event follows the updates in Eqs.~\eqref{eq:update_mass} and \eqref{eq:update}. Before every event we recompute the splitting and check the exterior, future-directed branch, source reachability, $D_1<0$, $D_3<0$, and $\mathcal E_2<0$. A proposed post-absorption state is also rejected if it has no horizon or has a smaller outer horizon. At fixed $l$ and logarithmic reference scale, differentiating $F(x_+;M,Q)=0$ gives the first law
\begin{equation*}
    dM=F'(x_+)\,dx_+-Q\ln x_+\,dQ=T_H\,dS_{\rm BH}+\Phi_+\,dQ,
\end{equation*}
where $T_H=F'(x_+)/(4\pi l)$, $S_{\rm BH}=4\pi l x_+$ in our $c=8G=1$ units, and $\Phi_+=-Q\ln x_+$. Here $F'$ denotes differentiation with respect to $x$. For infinitesimal absorption, $\delta M=E_2$ and $\delta Q=q_2$, so
\begin{equation*}
    T_H\,\delta S_{\rm BH}=F'(x_+)\,\delta x_+=E_2+q_2Q\ln x_+>0.
\end{equation*}
The last inequality follows from the future-directed condition at horizon crossing. Since $F'(x_+)>0$ for a nonextremal outer horizon, it implies $\delta x_+>0$ and hence irreducible-mass growth. The finite-step check tests the discrete update directly. A finite step can cross the marginal curve, the next event is then rejected, without claiming exact equality at the reported endpoint.

In the fixed reference convention we define~\cite{PhysRevLett.25.1596,PhysRevD.4.3552}
\begin{align}
    M_{{\rm irr},n}&=x_{+,n}^2,\\
    E_{{\rm able},n}&=M_n-M_{{\rm irr},n}=-\frac{Q_n^2}{2}\ln x_{+,n},\\
    E_{{\rm ed},n}&=M_0-M_n,\qquad \xi_n=\frac{E_{{\rm ed},n}}{n m_1\mathcal E_1},\\
    \Xi_n&=\frac{E_{{\rm ed},n}}{E_{{\rm able},0}-E_{{\rm able},n}}.
\end{align}
Here $M_{\rm irr}$ is the irreducible mass of a neutral black hole with the same horizon radius and hence the same entropy. $E_{\rm able}$ is the corresponding reversible mass-difference reservoir, and $\xi_n$ and $\Xi_n$ denote net energy return on investment (EROI) and energy utilization efficiency (EUE). The gross returned-to-injected energy ratio is $1+\xi_n$. The exact bookkeeping identity
\begin{equation}
    E_{{\rm able},0}-E_{{\rm able},n}=E_{{\rm ed},n}+M_{{\rm irr},n}-M_{{\rm irr},0}\label{eq:partition}
\end{equation}
shows that $\Xi_n$ and $1-\Xi_n$ are the shares of the consumed reservoir going into extracted energy and irreducible-mass growth respectively.
\begin{table*}[t]
    \centering
    \setlength{\tabcolsep}{5.5pt}
    \renewcommand{\arraystretch}{1.08}
    \caption{Case I: $M_0=0.5$, $Q_0=0.8$, $m_1=M_0/5000$, $x_{\rm d}=0.6$, $\mathcal E_1=2$, $\eta=0.78345$ and $\hat q_2=-2$. The incident charge is adaptive. After 2650 absorptions, the next event fails the incident turning-point condition. Irreducible-mass growth accounts for $31.23\%$ of the consumed reservoir. Rows show selected stages; dashes at $n=0$ indicate undefined efficiencies.}\label{tab:iterations1}
    \begin{tabular}{c c c c c c c c}
        \hline\hline
        $n$
        & $Q_n$
        & $M_n/M_0$
        & $E_{{\rm able},n}/M_0$
        & $E_{{\rm ed},n}/M_0$
        & $M_{{\rm irr},n}/M_0$
        & $\xi_n$
        & $\Xi_n$ \\
        \hline
        0 & 0.800000 & 1.000000 & 0.369306 & 0.000000 & 0.630694 & -- & --\\
        1 & 0.799888 & 0.999926 & 0.369173 & 0.000074 & 0.630753 & 0.186195 & 0.558175\\
        1325 & 0.651412 & 0.909829 & 0.228464 & 0.090171 & 0.681365 & 0.170133 & 0.640224\\
        2648 & 0.503047 & 0.836795 & 0.131962 & 0.163205 & 0.704833 & 0.154083 & 0.687628\\
        2649 & 0.502935 & 0.836746 & 0.131901 & 0.163254 & 0.704846 & 0.154071 & 0.687657\\
        2650 & 0.502823 & 0.836698 & 0.131840 & 0.163302 & 0.704858 & 0.154059 & 0.687686\\
        \hline\hline
    \end{tabular}
\end{table*}

\begin{table*}[t]
    \centering
    \setlength{\tabcolsep}{5.5pt}
    \renewcommand{\arraystretch}{1.08}
    \caption{Case II: $x_{\rm d}=0.85$, $\mathcal E_1=4$ and $\hat q_2=-5$; the other fixed data, including $m_1=M_0/5000$, are as in Table~\ref{tab:iterations1}. The last column gives the specific energy already absorbed in obtaining row $n$. Its final value is negative, but the next proposed value is $\mathcal E_{2,632}\simeq1.22\times10^{-4}>0$, so no 633rd event is accepted. Irreducible-mass growth accounts for $97.73\%$ of the consumed reservoir.}\label{tab:iterations2}
    \begin{tabular}{c c c c c c c c c}
        \hline\hline
        $n$
        & $Q_n$
        & $M_n/M_0$
        & $E_{{\rm able},n}/M_0$
        & $E_{{\rm ed},n}/M_0$
        & $M_{{\rm irr},n}/M_0$
        & $\xi_n$
        & $\Xi_n$
        & $\mathcal E_{2,n-1}$ \\
        \hline
        0 & 0.800000 & 1.000000 & 0.369306 & 0.000000 & 0.630694 & -- & -- & --\\
        1 & 0.799720 & 0.999986 & 0.368796 & 0.000014 & 0.631190 & 0.017682 & 0.027725 & -0.12614246\\
        316 & 0.711408 & 0.996645 & 0.248993 & 0.003355 & 0.747652 & 0.013272 & 0.027886 & -0.06321012\\
        630 & 0.623376 & 0.995527 & 0.172527 & 0.004473 & 0.823000 & 0.008875 & 0.022730 & -0.00047753\\
        631 & 0.623096 & 0.995527 & 0.172324 & 0.004473 & 0.823203 & 0.008861 & 0.022707 & -0.00027775\\
        632 & 0.622815 & 0.995527 & 0.172121 & 0.004473 & 0.823406 & 0.008847 & 0.022684 & -0.00007796\\
        \hline\hline
    \end{tabular}
\end{table*}

\begin{figure*}[t]
\centering\includegraphics[width=\textwidth]{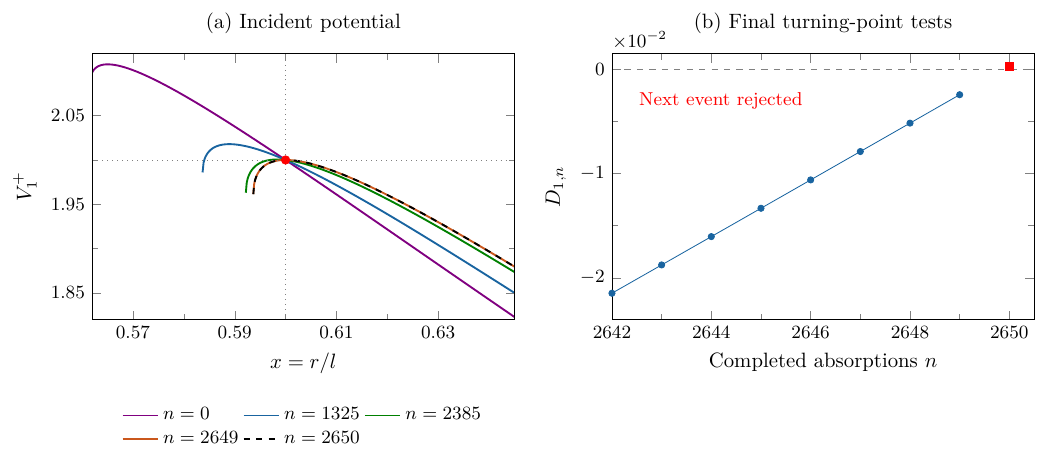}
    \caption{Case I of Table~\ref{tab:iterations1}. (a) Incident potentials for the updated $(M_n,Q_n,\hat q_{1,n})$, with each particle's charge fixed along its trajectory. All pass through $(x_{\rm d},\mathcal E_1)=(0.6,2)$ by construction. (b) The final slopes distinguish the last accepted pre-event state, $D_{1,2649}=-2.43817\times10^{-3}$, from the state after 2650 absorptions, $D_{1,2650}=2.90565\times10^{-4}$, which rejects the next event. The dashed curve in (a) corresponds to this next-event test.}\label{fig:eff_potential}
\end{figure*}

The examples in Tables~\ref{tab:iterations1} and \ref{tab:iterations2} use $k=M_0/m_1=5000$ and accept 2650 and 632 absorptions respectively. Mass-step refinement and higher-precision calculations support the stability of these results in Appendix~\ref{app:numerical_checks}. Tables~\ref{tab:iterations1} and \ref{tab:iterations2} illustrate the thermodynamic cost of repetitive extraction. Negative-energy absorption reduces the black-hole mass and charge while increasing its irreducible mass and entropy. According to Eq.~\eqref{eq:partition}, the consumed energy reservoir is divided between extracted energy and irreversible mass growth, and the much lower EUE in case II reflects the larger share in the latter. Case I terminates when the incident particle can no longer reach the prescribed decay point, whereas case II terminates when the next captured fragment no longer has negative energy. Both sequences stop with $E_{{\rm able},n}>0$: the protocol constraints fail before the reservoir is exhausted. Their nonzero residual charges realize the third-law analogy within the prescribed protocol.

\subsection{Potential evolution and parameter dependence}\label{sec:parameter_dependence}

\begin{figure*}[t]
\centering\includegraphics[width=\textwidth]{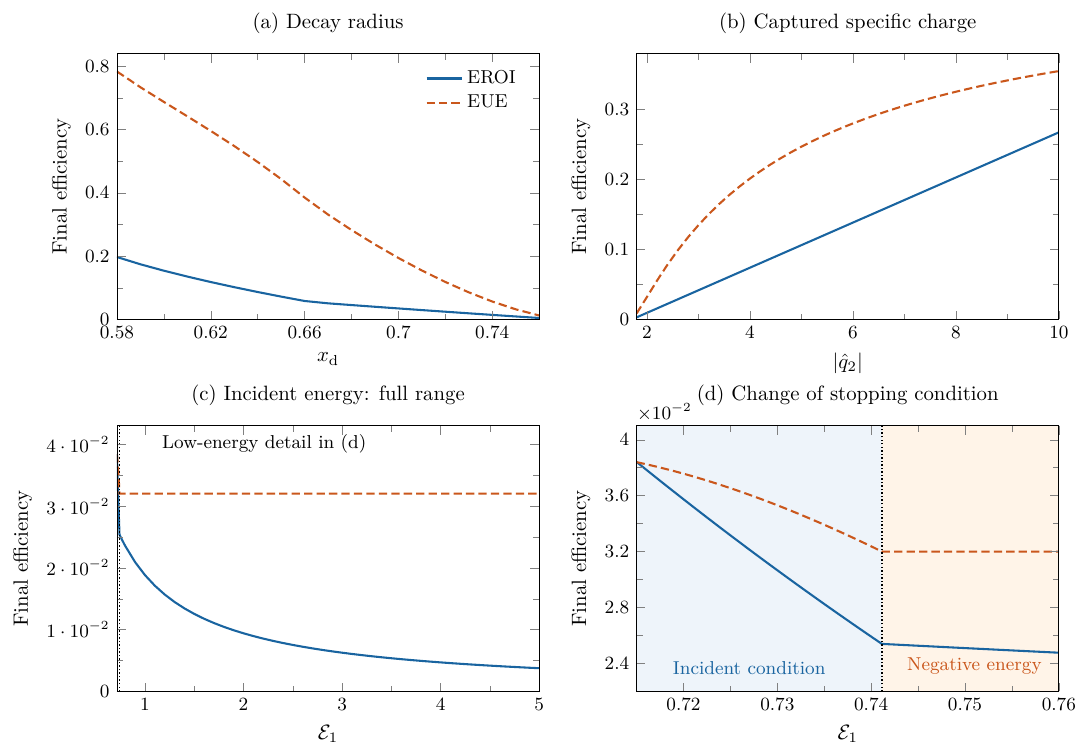}
    \caption{Final EROI (solid blue) and EUE (dashed orange) versus (a) decay radius, (b) captured specific charge and (c) incident energy; (d) enlarges the low-energy part of (c). Fixed inputs are $M_0=0.5$, $Q_0=0.8$, $m_1=M_0/50000$, $\eta=0.78345$, $x_{\rm d}=0.75$, $\hat q_2=-2$ and $\mathcal E_1=2$, with only the indicated quantity varied. In (d), shading distinguishes the incident-limited and negative-energy-limited regimes in the continuum limit; the dotted line marks the analytic switch $\mathcal E_{1,*}\simeq0.74114$.}\label{fig:efficiencies}
\end{figure*}

To illustrate how repetitive extraction changes the orbital kinematics, Fig.~\ref{fig:eff_potential} shows the evolution of the incident effective potential as the black-hole parameters and the incident charge are updated. In the illustrated case, the potential peak decreases and moves toward the fixed decay position, making the incident turning point increasingly marginal as $D_1$ approaches zero from below. A finite update changes $D_1$ to a positive value, making the immediately exterior region inaccessible at the prescribed energy, therefore the next proposed event is rejected.

To investigate how the extraction efficiency responds to different experimental configurations, Fig.~\ref{fig:efficiencies} compares the final EROI and EUE as functions of the decay radius, captured specific charge and incident specific energy. In panel (a), both efficiencies decrease as the decay radius increases over the displayed range. At fixed black-hole parameters, increasing the exterior decay radius makes the electrostatic contribution to $\mathcal E_2$ less negative and increases $\sqrt{F_{\rm d}}$, reducing access to negative-energy states. Panel (b) shows that increasing $|\hat q_2|$ raises both efficiencies in the studied range by driving the captured energy deeper below zero.

Panels (c,d) show that varying only the fixed incident energy switches the stopping condition. With the other inputs fixed, the captured energy and the updates in Eqs.~\eqref{eq:update_mass} and \eqref{eq:update} do not depend on $\mathcal E_1$. Different incident energies therefore share the same discrete mass-charge sequence, i.e., the successive black-hole states $(M_n,Q_n)$ after $n$ accepted absorptions, up to their respective stopping points. Changing $\mathcal E_1$ readjusts $\hat q_{1,n}$ through Eq.~\eqref{eq:adaptive_charge} and changes incident reachability. At lower incident energies in the displayed scan, the incident condition $D_1=V_1^{+\prime}(x_{\rm d})<0$, required for the particle to approach the decay point from larger radii, fails earlier and truncates the sequence. Increasing $\mathcal E_1$ extends the sequence toward $\mathcal E_2=0$. Later absorptions contribute less extracted energy while increasing irreducible mass, lowering EUE, EROI also decreases as the average gain per event falls and the injected energy rises. When incident and source reachability permit all negative-energy absorptions, the endpoint is common to those experiments: EUE reaches a plateau, while EROI decreases as $1/\mathcal E_1$ because the injected energy increases.

For the stated baseline, Appendix~\ref{app:continuum} gives the continuum switching energy \(\mathcal E_{1,*}\simeq0.741\). Under the stated applicability conditions, initially admissible sequences terminate through loss of incident reachability below this value and loss of negative-energy capture above it. The two limiting conditions are reached simultaneously at equality. The final efficiencies are therefore controlled jointly by incident reachability, negative-energy capture and the irreversible mass growth quantified by Eq.~\eqref{eq:partition}.

\section{Discussion}\label{sec:dis}

The logarithmic electrostatic potential and AdS confinement make it necessary to distinguish local splitting kinematics, globally allowed particle paths, and thermodynamic energy accounting. In the probe background, neutral test particles have no exterior circular geodesics. Charged circular motion requires $\mathcal Q\ge2\sqrt M$, subject to real angular momentum and a separate stability test. The single-event analysis gives a maximum energy at fixed daughter masses and charge allocation by optimizing the emission direction. The finite-mass example establishes negative-energy capture and collection at a finite receiver within the probe model for a locally prepared parent.

For the repetitive radial protocol, the incident specific charge is adjusted as the black-hole mass and charge evolve. The two examples stop when either the incident turning-point condition or the negative-energy capture condition fails. The continuum mass-charge path in Appendix~\ref{app:continuum} identifies how these mechanisms exchange priority as the fixed incident energy changes. Both sequences terminate with nonzero residual charge remaining extractable energy, $E_{{\rm able},n}>0$: the protocol constraints fail before the reservoir is exhausted. This provides a third-law analogy concerning incomplete depletion under fixed protocol constraints, but it does not establish the unattainability of zero temperature or a bound on the physical duration of the process.

Energy partition is also protocol dependent. The consumed energy reservoir is divided between extracted energy and irreversible growth of irreducible mass, with substantially different shares in the two examples. The residual reservoir and irreducible-mass growth represent distinct limitations on recovery of the initial extractable energy. Eq.~\eqref{eq:partition} shows that EUE measures the extracted share of the consumed reservoir, whereas the fraction of the initial reservoir recovered also depends on the energy remaining at termination. A higher EUE therefore need not imply a larger recovered fraction: comparisons between protocols must account for both irreversible mass growth and incomplete reservoir depletion. When incident and source reachability permit all negative-energy absorptions, the experiments at different incident energies share a common endpoint. EUE then reaches a plateau, while EROI decreases as $1/\mathcal E_1$.

These calculations include the electromagnetic field in each exact stationary metric but approximate the inter-stage backreaction by quasistatic updates. Mass-step refinement and higher-precision calculations support the stability of the reported results shown in Appendix~\ref{app:numerical_checks}. Radiation, self-force, a microscopic splitting model and preparation costs remain outside the present analysis.

\acknowledgments 
This work is supported by the National Natural Science Foundation of China (No. 12375059), the National Key Research and Development Program of China (Nos. 2021YFC2203001, 2021YFC2201901), and the Project of National Astronomical Observatories, Chinese Academy of Sciences (No. E4TG6601).

\appendix

\section{Continuum path, applicability, and finite steps}\label{app:continuum}

Let $a=-\hat q_2>0$, $\lambda=-\ln x_{\rm d}>0$, $S=\sqrt{F(x_{\rm d};M,Q)}$, and $h=m_1\tilde\mu_2$. The update reads $\Delta Q=-ha$ and $\Delta M=h(S-aQ\lambda)$. With $x_{\rm d}$ fixed and $S>0$, differentiating $S^2=x_{\rm d}^2-M+\lambda Q^2/2$ along the continuum limit of these updates gives
\begin{equation}
    \frac{dM}{dQ}=Q\lambda-\frac Sa,\qquad \frac{dS}{dQ}=\frac{1}{2a}.\label{eq:continuum_path1}
\end{equation}
The solution is
\begin{equation}
\begin{aligned}
    S(Q)&=b+\frac Q{2a},\\
    M(Q)&=x_{\rm d}^2+\frac{\lambda Q^2}{2}-\left(b+\frac Q{2a}\right)^2,
\end{aligned}
\label{eq:continuum_path}
\end{equation}
where $b=S_0-Q_0/(2a)$. This is the continuum limit of the specified recurrence, not a solution of time-dependent Einstein-Maxwell evolution. The fragment mass ratio cancels from the path but determines the finite step $h$.

Setting $\mathcal E_2=S-aQ\lambda=0$ on this path gives
\begin{equation}
    Q_E=\frac{b}{a\lambda-1/(2a)}.\label{eq:negative_endpoint}
\end{equation}
An accessible negative-energy endpoint requires $a\lambda-1/(2a)\ne0$, $0<Q_E<Q_0$, and $\mathcal E_2=S-aQ\lambda<0$ before $Q_E$. One must also retain $S>0$, a nonextremal outer horizon, $x_+<x_{\rm d}<1$, and the source condition $\mathcal E_1\ge\max\sqrt{1-M(Q)}$. The selected incident and outgoing conditions remain necessary. If any fails first, the expression for $Q_E$ alone does not describe the endpoint.

Introducing the incident threshold $B(M,Q)$, the incident condition in Eq.~\eqref{eq:turn_slope} takes the form
\begin{equation}
\begin{aligned}
    B(M,Q)&=S+\lambda\frac{x_{\rm d}^2-Q^2/4}{S},\\
    \mathcal E_1&>B(M,Q).
\end{aligned}
\label{eq:incident_threshold}
\end{equation}
This equivalence follows from $D_1=(\mathcal E_1-B)/(x_{\rm d}\ln x_{\rm d})$ and $x_{\rm d}\ln x_{\rm d}<0$. Along the continuum path, write $B(Q)\equiv B(M(Q),Q)$. For the switch to be given by $B(Q_E)$, the incident threshold must not have a larger interior maximum. Direct differentiation gives
\begin{equation}
\begin{aligned}
    2aS^2\frac{dB}{dQ}={} &b^2-\lambda x_{\rm d}^2+b(1/a-a\lambda)Q\\
    &+\frac14(1/a^2-\lambda)Q^2.
\end{aligned}
\label{eq:B_derivative}
\end{equation}
For $M_0=0.5$, $Q_0=0.8$, $x_{\rm d}=0.75$ and $a=2$, the right-hand side is a quadratic whose maximum over $Q_E\le Q\le Q_0$ is $-0.13647844$. Hence $B$ increases as charge decreases. Subject to the applicability conditions stated above, the switch between the two stopping mechanisms for this baseline occurs at
\begin{equation}
    \mathcal E_{1,*}=B(M(Q_E),Q_E).\label{eq:energy_switch}
\end{equation}
At $\mathcal E_1=\mathcal E_{1,*}$ the two limiting conditions, $D_1=0$ and $\mathcal E_2=0$, are reached simultaneously at $Q_E$. For $\mathcal E_1>\mathcal E_{1,*}$, the incident condition remains strictly satisfied down to $Q_E$, condition 2 therefore terminates the sequence first. For $\mathcal E_1<\mathcal E_{1,*}$, an initially admissible sequence reaches $D_1=0$ at a larger residual charge $Q>Q_E$, while $\mathcal E_2<0$; condition 1 therefore terminates the sequence first.

The finite-step lapse obeys the exact algebraic identity
\begin{equation}
    S_{n+1}^2=S_n^2-hS_n+\frac{\lambda a^2}{2}h^2.\label{eq:finite_S}
\end{equation}
These relations also distinguish the implications of the two negative-energy branches for repetitive extraction. Consider an initially admissible state satisfying $Q_0<Q_{\rm neg}(x_{\rm d};M_0)$, $x_e<x_{\rm d}<\sqrt{M_0}$ and $a<\sqrt{-1/\ln M_0}$, where the branch conditions are evaluated at the initial mass $M_0$. Its captured specific energy is $\mathcal E_{2,0}=S_0-a\lambda Q_0<0$. Since $x_e=\exp[-1/(2a^2)]$, the condition $x_{\rm d}>x_e$ is equivalent to $a^2\lambda<1/2$. Eqs.~\eqref{eq:continuum_path1} and~\eqref{eq:continuum_path} therefore yield
\begin{equation}
\begin{aligned}
    \mathcal E_2(Q) &=b+\left(\frac{1}{2a}-a\lambda\right)Q,\\
    \frac{d\mathcal E_2}{dQ} &=\frac{1}{2a}-a\lambda>0.
\end{aligned}
\label{eq:continuum_capture_energy}
\end{equation}
Because $Q$ decreases during extraction, it follows that
\begin{equation}
    \mathcal E_2(Q)-\mathcal E_{2,0}=\left(\frac{1}{2a}-a\lambda\right)(Q-Q_0)<0,\label{eq:negative_energy_monotonic}
\end{equation}
when $0<Q<Q_0$. Thus $\mathcal E_2$ remains strictly negative along the admissible continuum path, leaving the incident turning-point condition as the only limiting mechanism on this branch. More generally, if $a^2\lambda=1/2$, the captured energy is constant along the continuum path. 

For finite-step in the branch $a^2\lambda<1/2$, an extension follows directly from Eq.~\eqref{eq:finite_S}:
\begin{equation*}
    S_{n+1}^2-\left(S_n-\frac h2\right)^2=\frac{h^2}{2}\left(a^2\lambda-\frac12\right)<0.
\end{equation*}
If $0<h<2S_n$ and the proposed next state has $S_{n+1}>0$, the positive square root satisfies $S_{n+1}<S_n-h/2$. Using $Q_{n+1}=Q_n-ha$, we obtain
\begin{equation}
\begin{aligned}
    \mathcal E_{2,n+1}-\mathcal E_{2,n} &=S_{n+1}-S_n+a^2\lambda h \\ 
    &<h\left(a^2\lambda-\frac12\right)<0.
\end{aligned}
\label{eq:finite_negative_energy}
\end{equation}
Hence an initially negative captured energy remains negative at every successive step for which these conditions hold. Independently of the negative-energy branch, the two extraction conditions imply a positive charge bound for the prescribed protocol. For an admissible pre-event state, $x_{\rm d}>x_+\ge Q/2$ gives $x_{\rm d}^2-Q^2/4>0$. Negative-energy capture and incident reachability respectively require
\begin{equation*}
\begin{aligned}
    S&<a\lambda Q,\\\mathcal E_1
    &>S+\lambda\frac{x_{\rm d}^2-Q^2/4}{S} >\frac{x_{\rm d}^2-Q^2/4}{aQ}.
\end{aligned}
\end{equation*}
Consequently,
\begin{equation}
\begin{aligned}
    Q&>Q_{\rm lb},\\Q_{\rm lb}
    &\equiv 2\left[\sqrt{a^2\mathcal E_1^2+x_{\rm d}^2}-a\mathcal E_1 \right]>0.
\end{aligned}
\label{eq:protocol_charge_bound}
\end{equation}
For fixed finite $a$, $\mathcal E_1$ and $x_{\rm d}$, this bound excludes complete discharge along an admissible continuum sequence. It applies to pre-event states. At finite step $h$, the last accepted update instead gives $Q_{\rm end}>Q_{\rm lb}-ha$, thus $ha<Q_{\rm lb}$ is a sufficient condition for positive residual charge after that update. This condition is satisfied in both principal examples.

\section{Numerical stability and particle scales}\label{app:numerical_checks}

Over all accepted stages and all three particles, $\max_i|q_i|/Q_n$ is below $0.001710$ in case I and $0.005987$ in case II; $\max_i|E_i|/M_n$ is below $0.000537$ and $0.000815$ respectively. The maximum particle-mass ratios are below $0.000240$ and $0.000201$.

Table~\ref{tab:convergence} separates finite mass-step sensitivity from numerical precision. From $k=5000$ to $50000$, the residual charges change by $8.97\times10^{-5}$ and $1.40\times10^{-4}$ respectively, and the EUE values change by less than $6.25\times10^{-5}$ and $2.04\times10^{-5}$. A separate 40/60-digit calculation preserves both principal examples' accepted counts and displayed endpoint values. The Supplemental Material contains the scripts and machine-readable data for the tables and figures, including the parameter sampling, horizon-finding procedure, numerical precision and stopping criteria. For each sampled sequence, it also reports the maximum particle-scale ratios:
\begin{equation*}
\begin{aligned}
    \epsilon_q&=\max_{n,i}\frac{|q_{i,n}|}{Q_n},\\
    \epsilon_E&=\max_{n,i}\frac{|E_{i,n}|}{M_n},\\
    \epsilon_m&=\max_{n,i}\frac{m_i}{M_n},
\end{aligned}
\end{equation*}
where the maxima are taken over accepted pre-event stages and all three particles. These diagnostics cover the full parameter scans in Fig.~\ref{fig:efficiencies}.

%\vspace{5pt}
% Particle-scale checks also cover the full parameter range displayed in Fig.~\ref{fig:efficiencies}.

% Keep the final table in the text flow, before the entire bibliography.
\onecolumngrid
\vspace{5pt}
\newsavebox{\layoutTableIV}
\savebox{\layoutTableIV}{
\begin{minipage}{\textwidth}
\expandafter\def\csname @captype\endcsname{table}
\centering\small
\setlength{\tabcolsep}{4pt}
\renewcommand{\arraystretch}{1.0}
\caption{Mass-step refinement. Main Tables~\ref{tab:iterations1} and \ref{tab:iterations2} use $k=5000$. The coarse rows serve only to quantify discretization error. In particular, the $k=50$ examples are not uniformly within the test-particle regime.}\label{tab:convergence}
    \begin{tabular}{c|rrrr|rrrr}\hline\hline
& \multicolumn{4}{c|}{Case I: turning-point gate} & \multicolumn{4}{c}{Case II: negative-energy gate}\\
$k=M_0/m_1$ & $n_{\rm end}$ & $Q_{\rm end}$ & $M_{\rm end}$ & $\Xi_{\rm end}$ & $n_{\rm end}$ & $Q_{\rm end}$ & $M_{\rm end}$ & $\Xi_{\rm end}$\\\hline
50 & 28 & 0.48600185 & 0.41427836 & 0.69622385 & 7 & 0.60375116 & 0.49740771 & 0.02465531\\
100 & 54 & 0.49721607 & 0.41690991 & 0.69126392 & 13 & 0.61776893 & 0.49758991 & 0.02401750\\
500 & 266 & 0.50170176 & 0.41806428 & 0.68836743 & 64 & 0.62057249 & 0.49773222 & 0.02281686\\
1000 & 531 & 0.50226247 & 0.41820877 & 0.68800472 & 127 & 0.62197426 & 0.49774964 & 0.02275578\\
2000 & 1061 & 0.50254282 & 0.41828103 & 0.68782332 & 253 & 0.62267515 & 0.49775835 & 0.02272522\\
5000 & 2650 & 0.50282318 & 0.41834877 & 0.68768576 & 632 & 0.62281533 & 0.49776357 & 0.02268357\\
10000 & 5299 & 0.50287925 & 0.41836323 & 0.68764947 & 1263 & 0.62295551 & 0.49776531 & 0.02267747\\
50000 & 26492 & 0.50291289 & 0.41837235 & 0.68762330 & 6315 & 0.62295551 & 0.49776670 & 0.02266326\\
\hline\hline\end{tabular}
\end{minipage}%
}

%\newpage
% Reserve room for the whole table, the separator, and the start of the references.
\ifdim\dimexpr\ht\layoutTableIV+\dp\layoutTableIV+60pt\relax
  >\dimexpr\pagegoal-\pagetotal\relax
  \newpage
\fi
\noindent\usebox{\layoutTableIV}\par
\vspace{6pt}

% Draw the APS bibliography separator once, before returning to two columns.
\makeatletter
\begingroup
  \baselineskip14\p@
  \bib@device{\textwidth}{245.5\p@}%
\endgroup
\nobreak\@nobreaktrue\addvspace{4\p@}%
\makeatother
\par\twocolumngrid

% Keep individual references intact; column switching is handled above.
\makeatletter
\let\layout@NATbibliography\NAT@thebibliography
\def\NAT@thebibliography#1{%
  \layout@NATbibliography{#1}\interlinepenalty10000\relax
}
\renewcommand{\bibsection}{}
% Preserve the APS bibliography font and tighten only its line spacing.
\let\layout@bibfont\bibfont
\renewcommand{\bibfont}{\layout@bibfont\setlength{\baselineskip}{10pt}}
\makeatother
\bibliography{ref.bib}
\end{document}